\documentclass[longauthor]{aastex701}

\begin{document}

\title{A Census of Novae in the Small Magellanic Cloud, 1897--2026}

\author[0000-0002-1276-1486]{Allen W. Shafter}
\affiliation{Department of Astronomy and Mount Laguna Observatory, San Diego State University, San Diego, CA 92182, USA}
\email[show]{ashafter@sdsu.edu}

\begin{abstract}
A census of nova candidates in the Small Magellanic Cloud
(SMC) is presented.
Twenty-six optical transients have been reported as novae or nova-like
eruptions from the discovery of HV~849 in 1897 through 2026 October.
Twenty-three are probable classical novae. Two, MAXI~J0158$-$744
and CXOU~J005245.0$-$722844, are very fast, super-Eddington eruptions
that have been interpreted as novae on white dwarfs with Be-star companions, while a third,
SMCN~2002-10a, is a possible Be+WD event identified only from its quiescent
counterpart. Four additional Be+WD candidates that have been
detected only in X-rays are briefly discussed.
\end{abstract}


\section{Introduction}

The first nova recorded in the SMC, HV~849, was
identified as a variable star on Harvard plates by
H.~S.~Leavitt in 1897 but not
recognized as a nova until \citet{McKibben1951}, making it
the first extragalactic classical nova to be observed.
The SMC, with its younger and more metal-poor population compared with the Milky Way or M31, offers
an excellent opportunity to explore the dependence of nova properties
on the underlying stellar population.
Although the SMC is the second-nearest galaxy with a well-sampled nova population,
the existing compilations, the MPE online catalog of W.~Pietsch\footnote{\url{https://www.mpe.mpg.de/~m31novae/opt/smc/SMC_table.html}}
and \citet{Morel2021}, end in 2020, and the former contains errors that are
corrected here.
To facilitate future studies,
this Note provides an updated census.

\section{The Census}

Table~\ref{tab:novae} lists optical transients that
have been reported as novae or as nova-like eruptions on a white dwarf (WD)
within 5$^\circ$ of the SMC center (R.A.=00$^{\rm h}$52$^{\rm m}$45$^{\rm s}$,
Decl.=$-72^\circ49'43''$), including the Wing.
X-ray-selected Be+WD candidates without
associated optical outbursts are listed separately and are not counted
as novae; nor are AT2025yid (unclassified) and
two transients that have been discussed as possible novae:
ASASSN-16oh and AT2026ably. The former system reached only $V=16.8$ ($M_V\approx-2.2$),
and its spectra showed narrow disk emission lines with no sign of
ejecta.\footnote{Absolute magnitudes computed assuming $(m-M)_0 = 18.977$ \citep{Graczyk2020}.}
Its luminous supersoft X-ray emission has been attributed either to accretion
onto a massive WD without nuclear burning \citep{Maccarone2019} or
to a thermonuclear runaway without mass ejection
\citep{Hillman2019,Kato2020}. In either case it was not a classical nova.
The latter object reached $R=14.5$ ($M_R\approx-4.5$),
and faded by more than 4~mag within nine days.
Its low peak luminosity and amplitude suggest either a
cataclysmic variable \citep{Balam2026}, such as a foreground dwarf nova,
or an ASASSN-16oh-like event in the SMC.

Data through 2020 are from the MPE catalog, except where
noted in Table~\ref{tab:novae}. Only four systems have known
spectroscopic classes (two Fe\,II, one He/N, one hybrid) and fewer than half have
decline times, so the SMC sample remains poorly characterized relative
to the LMC and M31 \citep{Shafter2026a,Shafter2026d}.

\section{Be+WD Systems}

The recent recognition that the M31 recurrent nova 2017-01e likely arose
from a Be+WD progenitor \citep{Chamoli2025} raises the possibility
that such systems may represent a hitherto unappreciated population of novae.
Despite its relatively low mass,
the SMC, with its recent star formation and young stellar population,
hosts most of the known Be+WD candidates. Two SMC events have exhibited the fast,
luminous, supersoft eruptions expected from novae on massive WDs in
such systems. MAXI~J0158$-$744 (SMCN~2011-11a) showed a
very fast super-Eddington X-ray outburst with a brief optical brightening
\citep{Li2012}, while CXOU~J005245.0$-$722844 (SMCN~2024-05a) produced a
super-Eddington supersoft outburst lasting $<16$~d, accompanied by an optical
brightening of $\sim$0.5~mag that lasted $<7$~d. Its OBe companion was
subsequently confirmed spectroscopically \citep{Gaudin2024}.

SMCN~2002-10a \citep{Liller2002}, which lacks eruption spectroscopy or X-ray data,
is a weaker case. The quiescent counterpart
has the colors of a Be star that shows irregular ($\sim$0.1~mag) and
$\sim$30-d periodic ($\sim$0.05~mag) variability, prompting \citet{Mroz2016a}
to exclude it from their nova list as a possible low-amplitude Be-star
outburst. It is included here as a possible Be+WD nova.

Four additional SMC Be+WD candidates were found as supersoft X-ray sources with
Be-star counterparts but none have observed optical outbursts:
Swift~J011511.0$-$725611 \citep{Kennea2021}, Swift~J004427.3$-$734801
\citep{Coe2020}, XMMU~J010147.5$-$715550 \citep{Sturm2012},
and Suzaku~J0105$-$72. The latter,
discovered through a short-lived supersoft flash in 2005 August \citep{Takei2008},
now has a confirmed Be companion \citep{Cracco2018}, suggesting
the flash may have been the signature of an unobserved nova.

\begin{acknowledgments}
I thank Mati Morel for corrections to several nova positions and dates.
Claude (Anthropic) assisted with editing and with cross-checking the table;
the author takes full responsibility for the content.
\end{acknowledgments}

\setlength{\tabcolsep}{3.5pt}
\begin{deluxetable*}{ccccccccc}
\tabletypesize{\footnotesize}
\tablecolumns{9}
\tablewidth{0pt}
\tablecaption{Nova Candidates in the Small Magellanic Cloud\label{tab:novae}}
\tablehead{
\colhead{Nova} & \colhead{JD} & \colhead{RA} & \colhead{Decl.} & \colhead{Max} & \colhead{Min} &
\colhead{$t_2$} & \colhead{References} & \colhead{Comments} \\
\colhead{(SMCN)} & \colhead{($-$2\,400\,000)} & \colhead{(deg)} & \colhead{(deg)} & \colhead{(mag)} & \colhead{(mag)} &
\colhead{(d)} & \colhead{} & \colhead{}
}
\startdata
\cutinhead{Optical eruptions (the 26 counted events)}
1897-10a & 14213.0 & 15.5475 & $-$70.0489 & 11.4 pg & \nodata & \nodata & 1,2,3,4 & HV 849; pos.: ref.\ 4 \\
1927-10a & 25157.5 & 8.5039 & $-$73.2686 & 11.4 pg & \nodata & \nodata & 1,2,4 & pos.: ref.\ 4 \\
1951-06a\tablenotemark{a} & 33858.0 & 8.8429 & $-$72.9699 & 11.8: R & \nodata & $\sim$20 & 1,2,4 & Fe\,II?; pos.: ref.\ 4 \\
1952-12a & 34368.5 & 12.1220 & $-$73.5129 & 11.0 pg & \nodata & \nodata & 1,4,5 & pos.: ref.\ 4 \\
1974-08a & 42278.5 & 6.5088 & $-$74.0181 & 11.0 V & \nodata & \nodata & 1,4 & pos.: ref.\ 4 \\
1986-10a & 46709.0 & 9.2186 & $-$72.0780 & 10.1 V & \nodata & \nodata & 1,4 & pos.: ref.\ 4 \\
1994-06a & 49504.5 & 12.8737 & $-$73.3277 & 14.5 B & \nodata & 33 & 1 & \\
1996-06a & 50239.3 & 13.6793 & $-$72.5391 & 13.5 V & \nodata & \nodata & 1 & \\
1998-12a & 51173.0 & 15.5908 & $-$73.3075 & 13.5 V & \nodata & \nodata & 1 & \\
1999-07a & 51387.6 & 13.4167 & $-$72.2089 & 13.6 R & \nodata & 16 & 1 & \\
1999-08a & 51393.5 & 12.2154 & $-$72.5042 & 13.0 V & \nodata & \nodata & 1 & \\
1999-09a & 51439.7 & 14.8458 & $-$73.1322 & 12.8 R & \nodata & \nodata & 1 & \\
2001-10a & 52191.7 & 11.6163 & $-$73.4960 & $<12.1$ V & $\sim$20.0 I & 40 & 1,6,7 & Fe\,II; max.\ missed \\
2002-10a & 52563.7 & 14.1268 & $-$72.6079 & 12.3 V & 15.7 I & \nodata & 1,6,8 & Be+WD? \\
2004-06a & 53169.5 & 13.7634 & $-$73.6343 & 14.2 I & $\sim$19.1 I & \nodata & 1,6 & date: ref.\ 6; rebrightening \\
2005-08a & 53588.9 & 18.7503 & $-$73.4271 & 11.6 V & $>21.5$ I & 14 & 1,6 & Fe\,II \\
2006-08a & 53950.5 & 14.2661 & $-$73.1699 & $<11.8$ I & $>21$ I & 31.5 & 1,6 & date: ref.\ 6; max.\ missed \\
2008-10a & 54770.7 & 13.9072 & $-$72.0540 & 13.0 I & $\sim$20.8 I & 170 & 1,6 & \\
2011-11a & 55876.7 & 29.8578 & $-$74.2578 & $<14.3$ I & 14.8 I & \nodata & 1,6,9 & MAXI J0158$-$744; Be+WD \\
2012-06a & 56083.5 & 8.2294 & $-$74.3388 & 11.0 I & \nodata & 40 & 1,6 & O--Ne \\
2012-09a & 56187.7 & 12.0303 & $-$72.7761 & 12.1 I & $\sim$20.2 I & 7 & 1,6 & \\
2016-10a & 57670.7 & 16.5136 & $-$74.7877 & 8.5 w & 20.7 V & 4.0 & 1,10,11,12 & He/N; Galactic? (ref.\ 11) \\
2019-07a & 58668.8 & 16.1258 & $-$73.7730 & $<11.0$ g & 20.7 V & \nodata & 1,13,14,15 & hybrid \\
2020-10a & 59148.5 & 12.1857 & $-$72.8983 & 10.6 V & \nodata & \nodata & 1,14 & max.: ref.\ 14 \\
2024-05a & 60456.9 & 13.1879 & $-$72.4788 & $\sim$14.4 I & 14.9 V & \nodata & 16 & CXOU J0052$-$7228; Be+WD \\
2026-06a & 61202.9 & 12.6881 & $-$71.6436 & 10.5 g & 19.6 UVW1 & 6 & 17,18,19 & ASASSN-26dt; CSM; SSS \\
\cutinhead{Supersoft X-ray Be+WD candidates (no optical outburst; not counted)}
AzV 281 & 51651.4 & 15.4479 & $-$71.9306 & \nodata & 14.5 V & \nodata & 20 & XMMU J0101; nova unlikely \\
Swift J0115 & $\sim$59212 & 18.7958 & $-$72.9364 & \nodata & $\sim$14.4 V & \nodata & 21 & $\sim$3-mo.\ outburst; $P\approx17.4$~d \\
Swift J0044 & 58870 & 11.1138 & $-$73.8003 & \nodata & 15.0 V & \nodata & 22 & $\sim$21.5-d optical flares \\
Suzaku J0105 & 53613 & 16.4542 & $-$72.1239 & \nodata & 14.6 B & \nodata & 23,24 & single flash; pos.\ $\pm$20$''$ \\
\enddata
\tablenotetext{a}{Should be SMCN 1951-07a; name retained from earlier catalogs.}
\tablecomments{J2000 coordinates; JD = discovery date; magnitudes as reported
(pg = photographic; w = unfiltered). Min: quiescent magnitude from the listed
references.}
\tablerefs{(1) \url{https://www.mpe.mpg.de/~m31novae/opt/smc/SMC_table.html}; (2) \citet{Henize1954}; (3) \citet{McKibben1951}; (4) \citet{Morel2021}; (5) \citet{Smith1954}; (6) \citet{Mroz2016a}; (7) \citet{Mason2005}; (8) \citet{Liller2002}; (9) \citet{Li2012}; (10) \citet{Aydi2018}; (11) \citet{Hachisu2018}; (12) \citet{Shumkov2016}; (13) \citet{Aydi2019ATel12907}; (14) \citet{Watson2006}; (15) \citet{Izzo2022}; (16) \citet{Gaudin2024}; (17) \citet{Chomiuk2026}; (18) \citet{Kuin2026}; (19) \citet{Gaudin2026}; (20) \citet{Sturm2012}; (21) \citet{Kennea2021}; (22) \citet{Coe2020}; (23) \citet{Takei2008}; (24) \citet{Cracco2018}.}
\end{deluxetable*}

\bibliography{smc_novae}{}
\bibliographystyle{aasjournalv7}

\end{document}